\documentclass[aps,twocolumn,showpacs,preprintnumbers,amsmath,amssymb,floatfix]{revtex4}

\usepackage{graphicx}
\usepackage{dcolumn}
\usepackage{bm}
\usepackage[german,american]{babel}

\begin{document}


\title{The Origin of the Decoupling of Oxygen and Silicon Dynamics in Liquid Silica as Expressed by its Potential Energy Landscape}

\author{A. Saksaengwijit}
\author{A. Heuer}%
\affiliation{
Westf\"{a}lische Wilhelms-Universit\"{a}t M\"{u}nster, Institut f\"{u}r Physikalische Chemie\\
and International Graduate School of Chemistry\\
Corrensstr. 30, 48149 M\"{u}nster, Germany }

\date{\today}

\begin{abstract}
The oxygen and silicon dynamics in silica is compared via computer
simulations. In agreement with experimental data and previous
simulations a decoupling of oxygen and silicon dynamics is observed
upon cooling. The origin of this decoupling is studied in the
framework of the potential energy landscape. From analysis of the
transition features between neighboring superstructures of minima,
denoted metabasins, the differences between the oxygen and the
silicon dynamics can be quantified. The decoupling can be explicitly
related to the presence of generalized rotational processes, giving
rise to oxygen but not to silicon displacement. Closer analysis of
these processes yields important insight into the nature of the
potential energy landscape of silica. The physical picture of
relaxation processes in silica, obtained in previous work for the
oxygen dynamics, is consistent with the decoupling effects,
elucidated here.

\end{abstract}

\maketitle
\section{Introduction}
Silica is a well-known prototype network glass former. It is
classified as a strong liquid in the Angell scheme
\cite{Angell:1991} because the temperature dependence of its
dynamics displays an Arrhenius behavior. Interestingly, oxygen and
silicon dynamics are characterized by different activation energies.
Close to the glass transition temperature $T_g$ the difference
between the self-diffusion constants for oxygen $D_O$ and silicon
$D_{Si}$ is nearly two orders of magnitude
\cite{Mikkelsen:1984,Brebec:1980}. Moreover, it has been shown that
the viscous dynamics of liquid silica close to $T_g$ is strongly
controlled by the silicon dynamics \cite{Nascimento:2006}. The large
difference in diffusivity indicates different relaxation mechanisms
for oxygen and silicon atoms in liquid silica at low temperatures.

The decoupling of silicon and oxygen single-particle dynamics has
been already studied via computer simulations of SiO$_2$
\cite{Horbach:1999} modelled by the BKS potential \cite{BKS:1990}.
It turned out that in the limit of very high temperatures the
diffusive dynamics of oxygen and silicon is similar ($D_O/D_{Si}
\approx 1.3$). Upon cooling the slowdown in silicon dynamics is
significantly stronger as compared to that in oxygen dynamics. This
results in a faster decrease of $D_{Si}$ at low temperatures, e.g at
2800K one has $D_O/D_{Si} \approx 2.3$ \cite{Horbach:1999}. This has
been regarded as an evidence for a {\it decoupling dynamics}.
Interestingly, the temperature dependence of the average time-scale
for the lifetime of Si-O bonds follows $D_O(T)$ rather than
$D_{Si}(T)$ \cite{Horbach:1999}. Furthermore, the self part of the
van Hove correlation function for oxygen displays a clear additional
peak, which corresponds to the hopping dynamics. In contrast, no
clear hopping peak is present for the case of silicon
\cite{Horbach:1999}. Moreover, the recent study on the facilitated
dynamics in liquid silica \cite{Vogel:2004, Teboul:2004} shows that
the cooperative motion for oxygen dynamics can be described by a
string-like motion \cite{Donati:1998} whereas the usual string-like
motion is absent for the case of silicon atoms.

The potential energy landscape (PEL) has been shown to be an
indispensable tool to elucidate the nature of the dynamics and
thermodynamics of liquids \cite{Wales:2003book,Shell:2002,
Middleton:2001, Doliwa_Hop:2003,Saika:2004,Voivod:2004,
Stillinger:1982,Sastry:1998,Sciortino:2005}. In this framework the
dynamics of the system is represented by the dynamics of a point in
the high dimensional configurational space \cite{Goldstein:1969}.
The thermodynamics of the system can be expressed in terms of the
statistical properties of the inherent structures (IS), i.e. the
local minima of the PEL \cite{Stillinger:1982, Sciortino:2005,
LaNave:2002}. Naturally, the essential features observed in the
small subsystem are trivially averaged out on a macroscopic level
\cite{Doliwa_finite:2003,Saksaengwijit:2004}. Therefore, to
understand the decoupling dynamics in terms of the PEL properties it
is important to analyze sufficiently small systems without relevant
finite-size effects.

Recent studies on the {\it thermodynamics} of this model have shown
that the distribution of IS energies $G(e)$ possesses a low-energy
cutoff $e_c$ \cite{Saksaengwijit:2004}. Below $e_c$ the
configurational energy is strongly depleted as compared to the
Gaussian distribution, which describes $G(e)$ for $e > e_c$. An
important step to characterize the {\it dynamics} has been the
introduction of metabasins (MB) \cite{Doliwa_Hop:2003,
Saksaengwijit:2006}. This corresponds to a coarse-graining of the
configuration space. Groups of minima between which the system
performs forward and backward jumps are regarded as single states,
i.e MB. The total dynamics can then be formally described as a
sequence of jumps between MB, each characterized by a waiting time
$\tau$ and an energy $e$ which is chosen as the minimum IS energy of
the corresponding group of IS. Actually, the distribution $G(e)$ of
MB energies is virtually identical to that of IS such that the
thermodynamics does not change when switching from the IS to the MB
description. It turns out for silica that on this level of
description the oxygen dynamics of the system can be (to a good
approximation) characterized as a random-walk in configuration space
with a temperature-independent average hopping distance
\cite{Saksaengwijit:2006}.  This provides a strong link between
long-range transport and microscopic dynamics. As an immediate
consequence the temperature dependence of $D_O(T)$ is determined by
the temperature dependence of the average waiting time. Furthermore,
it has been shown that the diffusion and relaxation in liquid silica
has been related to the defect structure \cite{Martin-Samos2:2005,
Mousseau:2000, Saksaengwijit:2006} since close to defects the
mobility is largely enhanced.

The goal of this paper is twofold. First, we want to show why the
decoupling of oxygen and silicon dynamics as well as the
proportionality of the oxygen diffusion with the inverse bond
lifetime is a natural consequence of the nature of the dynamics of
BKS-silica. It will turn out that it is the presence of rotational
displacements which are the origin of the decoupling. Second, we
will relate the occurrence of these specific displacements to the
properties of the PEL. This will be another example where the
reference to properties of the PEL gives new insight into the
complex dynamics of supercooled liquids.

The outline is as follows. In Sect.II we present some computational
details. In Sect.III we show a detailed analysis of the nature of
the MB transitions. In this way the relevance of rotational
processes is elucidated. Based on these results we can unambiguously
explain the decoupling of the oxygen and silicon diffusion in terms
of these processes, as shown in Sect.IV. In Sect.V the rotational
processes are related to properties of the PEL. We close with a
discussion and a summary in Sect.VI. In particular, we show that the
general physical picture of the relaxation processes in silica, as
derived in our previous work \cite{Saksaengwijit:2006}, is fully
compatible with the results, obtained for the relation of the oxygen
and silicon dynamics in this work.

\section{Technical Details}
We have performed molecular dynamics simulations of liquid silica at
constant volume and energy (microcanonical ensemble). The density of
liquid silica is $\rho=2.30$ gcm$^{-3}$, which is similar to that in
previous simulations \cite{Voivod:2004,Horbach:1999}.  The standard
BKS potential \cite{BKS:1990} is used to describe the interaction
energy for liquid silica. The Coulombic interaction is calculated
with the Ewald-summation technique. The short range interaction is
truncated and shifted to zero at the cutoff radius $r_c$. To avoid
an energy drift $r_c$ is set to 8.5 \AA, which is slightly larger
than the one used previously \cite{Horbach:1999}.  In this work we
analyze the temperature range from 2800K to 4000K, using
well-equilibrated samples.  All results have been averaged over
three independent runs in order to gain better statistics.

It has been shown that for BKS-silica a system size of approx. 100
particles with periodic boundary conditions is sufficient to avoid
relevant finite-size effects for the configurational entropy, the
relaxation dynamics of BKS-silica (in terms of the diffusive
activation energy and the degree of non-exponentiality in the
incoherent scattering function) in the accessible temperature range
and the properties of tunneling systems
\cite{Saksaengwijit:2004,Saksaengwijit:2006,Martin-Samos2:2005,Reinisch:2005}.
As in our previous work we use a system size of $N=99$. To
characterize the distance between two configurations we use the
displacement of the oxygen atoms if not mentioned otherwise.

For each equilibrium run, the system is quenched via the standard
conjugated-gradient method. The times for the quench have been
optimized via the bisectioning method to enable an efficient search
for the precise location of the metabasin transitions
\cite{Doliwa_Hop:2003}. In this way we obtain a time series of IS
energies, which contains all relevant IS transitions which are
necessary to characterize all MB transitions. As mentioned above MB
are characterized by a waiting time $\tau$ and an energy $e$. The
energy of a MB is defined as the lowest IS energy in a MB (see
\cite{DoliwaEacc:2003} for a precise definition of MB).

For later purposes we introduce two different energy distributions,
relevant at temperature $T (=1/(k_B \beta))$.
\begin{equation}
p(e,T) \propto G(e) \exp(-\beta e)
\end{equation}
is the Boltzmann weighted distribution of MB energies at temperature
$T$ \cite{comment1}. It expresses the probability that at a randomly
given time the system is in a MB of energy $e$. Furthermore, we
define
\begin{equation}
\label{eqdefvarphi}
 \varphi(e,T) = \frac{p(e,T)\langle \tau(T)
\rangle}{\langle \tau(e,T) \rangle}
\end{equation}
where $\langle \tau(e,T)\rangle$ is the average waiting time of a MB
with energy $e$. $\varphi(e,T)$ denotes the probability that after a
MB transition the new MB has the energy $e$ \cite{Doliwa_Hop:2003}.

\section{Dynamic properties of oxygen and silicon atoms}

\subsection{Metabasin analysis}

In the previous studies on the binary Lennard-Jones system (BMLJ) it
has been shown that the mean square displacement  $R^2(n_{MB})$
after $n_{MB}$ transitions between MB is independent of temperature
except for extremely high temperatures for which minima of the PEL
are not relevant due to the dominance of anharmonic effects
\cite{Doliwa_Hop:2003}. As a consequence the diffusion constant
$D(T)$ is related to the average MB waiting time $\langle \tau(T)
\rangle$ via a temperature-independent jump length $ d$, i.e.
\begin{equation}
\label{eqdtau}
 D(T)= d^2/6N\langle \tau(T) \rangle.
\end{equation}

\begin{figure}
\includegraphics[width=8.6cm]{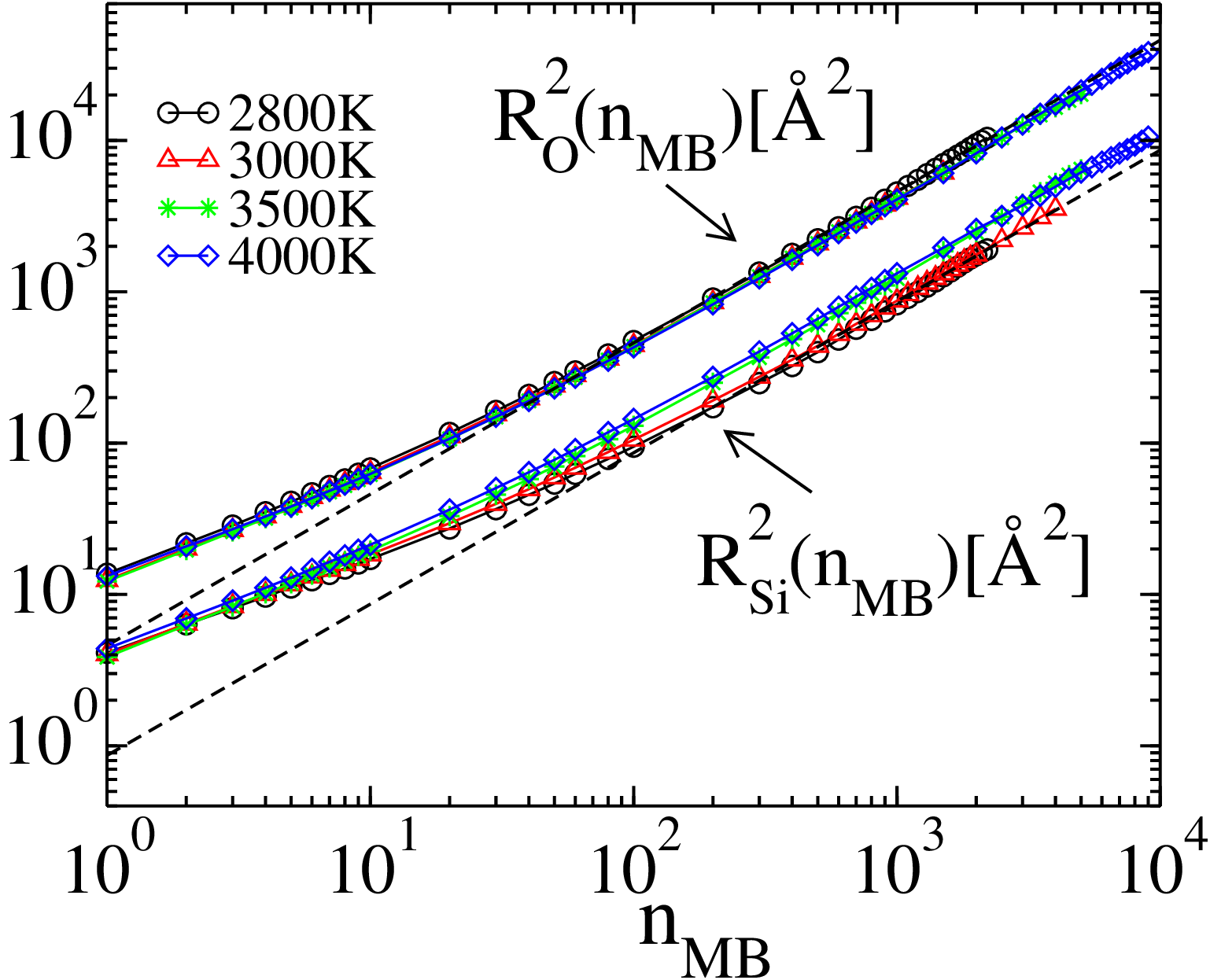}
\caption{\label{osimbr2}The average mean square displacement of
oxygen and silicon after $n$ MB transitions, i.e. $R^2_{O}(n_{MB})$
and $R^2_{Si}(n_{MB})$. The value of $n_{MB}$ indicates the number
of MB transitions. For the weak but significant temperature
dependence of $R^2_{Si}(n_{MB})/n_{MB}$ in the constant regime for
large $n_{MB}$ one obtains 1.3 \AA$^2$, 1.2 \AA$^2$, 0.9 \AA$^2$ and
0.8 \AA$^2$ for the four temperatures (starting from high
temperatures), respectively. }
\end{figure}

Here we repeat this analysis for the oxygen and silicon dynamics in
silica by calculating $R^2_{O}(n_{MB})$ and $R^2_{Si}(n_{MB})$,
respectively; see Fig.\ref{osimbr2}. One directly observes that
$R^2_{O}(n_{MB})$ is temperature-independent whereas there is a
significant and systematic temperature shift of $R^2_{Si}(n_{MB})$.
Actually, the temperature independence of  $R^2_{O}(n_{MB})$ has
been already used in our previous work \cite{Saksaengwijit:2006} to
justify the use of Eq.\ref{eqdtau} also for silica. Of course, the
decoupling of oxygen and silicon directly follows from the different
behavior of $R^2_{O}(n_{MB})$ and $R^2_{Si}(n_{MB})$.

On a qualitative level these observations agree with the previous
analysis of the decoupling of oxygen and silicon
\cite{Horbach:1999}. By analyzing the time decay $\tau_B$ of the
Si-O bond residual time, Horbach and Kob observed that $\tau_B \cdot
D_O$ does not depend on temperature whereas $\tau_B \cdot D_{Si}$
shows a significant temperature dependence at low temperatures.
$\tau_B$ should be proportional to the waiting time of MB
$\tau_{MB}$ because typical MB transitions will involve
bond-breaking events \cite{Saksaengwijit:2006}. Thus the previous
observations $D_O^{-1} \propto \tau_{B}$ but $D_{Si}^{-1} \not
\propto \tau_{B}$ are indeed qualitatively equivalent to the results
in Fig.\ref{osimbr2}.

But how to rationalize the temperature dependence of
$R^2_{Si}(n_{MB})$? The most simple explanation would be that
typical silicon displacements during MB transitions are smaller at
lower temperatures. In this case the temperature dependence would
reflect a change of the jump mechanism with temperature. In
literature, the complicated relaxation of silicon atoms has been
related to the possible presence of deep potential traps for silicon
\cite{Horbach:1999}. In this scenario, however, one would expect
that there is also a systematic temperature dependence for, e.g.
$R^2_{Si}(n_{MB}=1)$, which corresponds to the typical displacement
after one MB transition. Interestingly, in our simulations we have
not found such a temperature dependence; see Fig.\ref{osimbr2}.

\subsection{Rotational and translational periods}

Here we show that a different scenario is responsible for the
observations, shown in Fig.\ref{osimbr2}. For each time series of IS
(MB) at different temperatures, we have observed some rotational
processes. More precisely this means that during the time evolution
the system is found in different configurations which are identical
except for a permutation of oxygen atoms. The most simple
realization of rotational processes are rotations of a SiO$_4$
tetrahedral unit along the three-fold (C$_3$) or two-fold (C$_2$)
axis but also more complex permutations have been observed.

To elucidate this possibility in a quantitative way we first define
{\it rotational periods} as the maximum time intervals $[t_1,t_2]$
with the property that the configurations at $t_1$ and $t_2$ are
identical except for the permutation of oxygen atoms. Of course,
during one rotational period several rotations may take place. To
characterize the dynamics during a rotational period further one can
identify the different oxygen permutations. They are denoted {\it
rotational states}. Naturally, each rotational period contains at
least two rotational states but also more may be present.

\begin{figure}
\includegraphics[width=8.6cm]{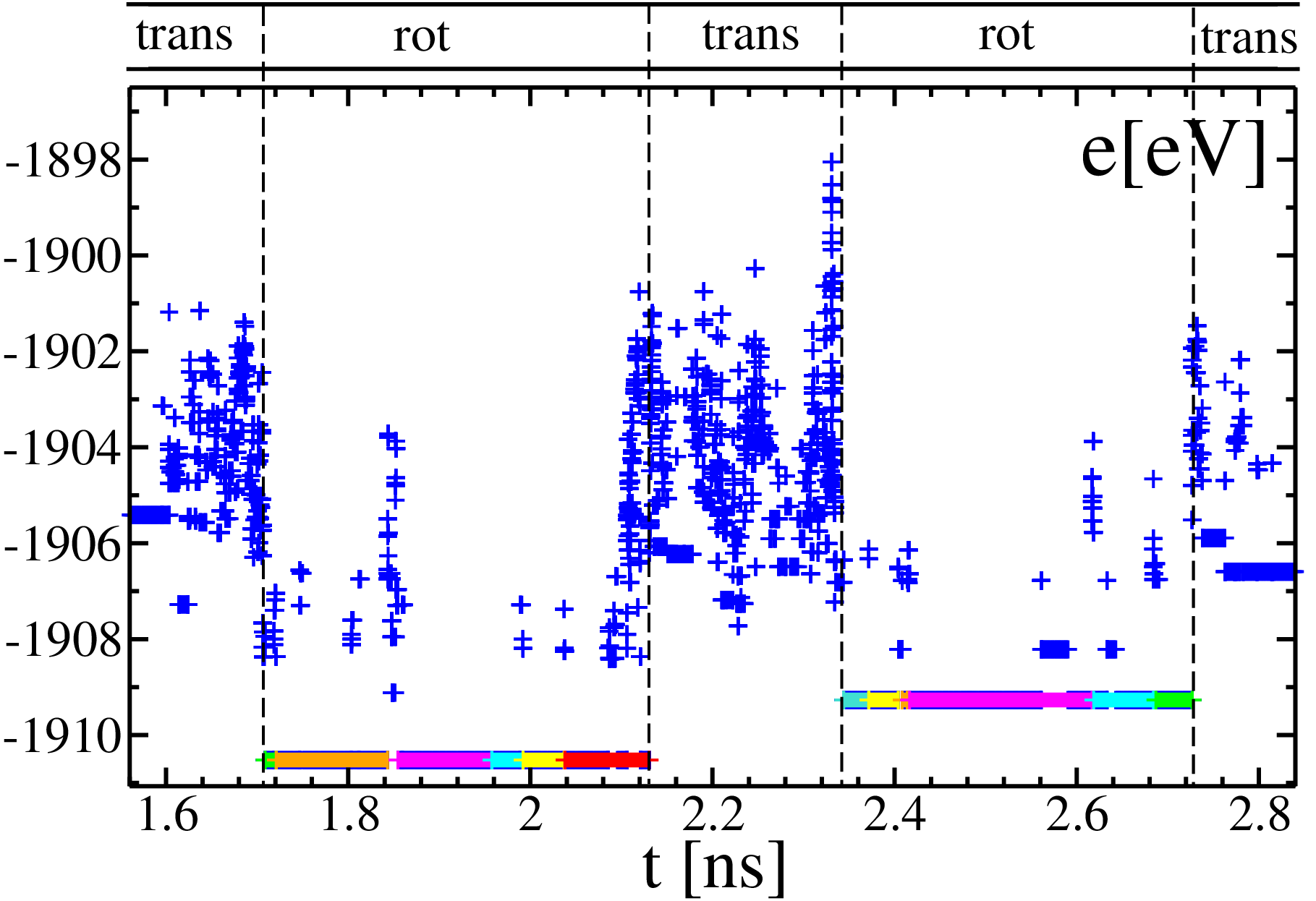}
 \caption{\label{timeseries}The time series of IS energies at 3500K.
The translational and rotational periods are labeled by {\bf trans}
and {\bf rot}, respectively.}
\end{figure}

In Fig.\ref{timeseries} a sequence of IS is shown. The different
rotational states which are visited during this time period are
indicated. Note that within the rotational period the system may
escape from a rotational state and either come back to the same
state or end up in another permutation of oxygen atoms. Typically,
subsequent rotational states are not connected by a single saddle
but rather by high-energy IS. We note in passing that for all
simulation runs we have not observed any permutation of silicon
atoms. The time intervals between the rotational periods are denoted
{\it translational periods}; see Fig.\ref{timeseries}. Whereas the
contributions from the long-time dynamics of oxygen atoms come from
rotational and translational periods, the long-time dynamics of
silicon atoms is exclusively related to the translational periods.

On average, many IS transitions are required for the system to go
from one rotational state to another; see Fig.\ref{timeseries}.
During this time also the silicon atoms display significant motion
but finally go back to the original position. From a single-atom
perspective during rotational processes the position of some oxygen
atoms change. Thus, rotational processes contribute to the
long-range dynamics of oxygen as opposed to that of silicon. The
simulation data show that rotational periods become more relevant at
low temperatures (see below for a quantification of this statement).
Thus, the fraction of MB transitions which finally are irrelevant
for the silicon dynamics increases with decreasing temperature. From
this observation it is indeed expected that $R^2_{Si}(n_{MB})$ is
smaller at lower temperatures as observed in Fig.\ref{osimbr2}.
Furthermore, a much weaker temperature dependence should occur for
$R^2_{Si}(n_{MB}=1)$ because on the single-transition scale a
silicon atom hardly realizes that a few MB-transitions later it will
go back to the original configuration. Stated differently, the
strong forward-backward correlations of the silicon dynamics during
the rotational periods should show up by a stronger subdiffusive
behavior in $R^2_{Si}(n_{MB})$ in agreement with Fig.\ref{osimbr2}.

\begin{figure}
\center
\includegraphics[width=8.6cm]{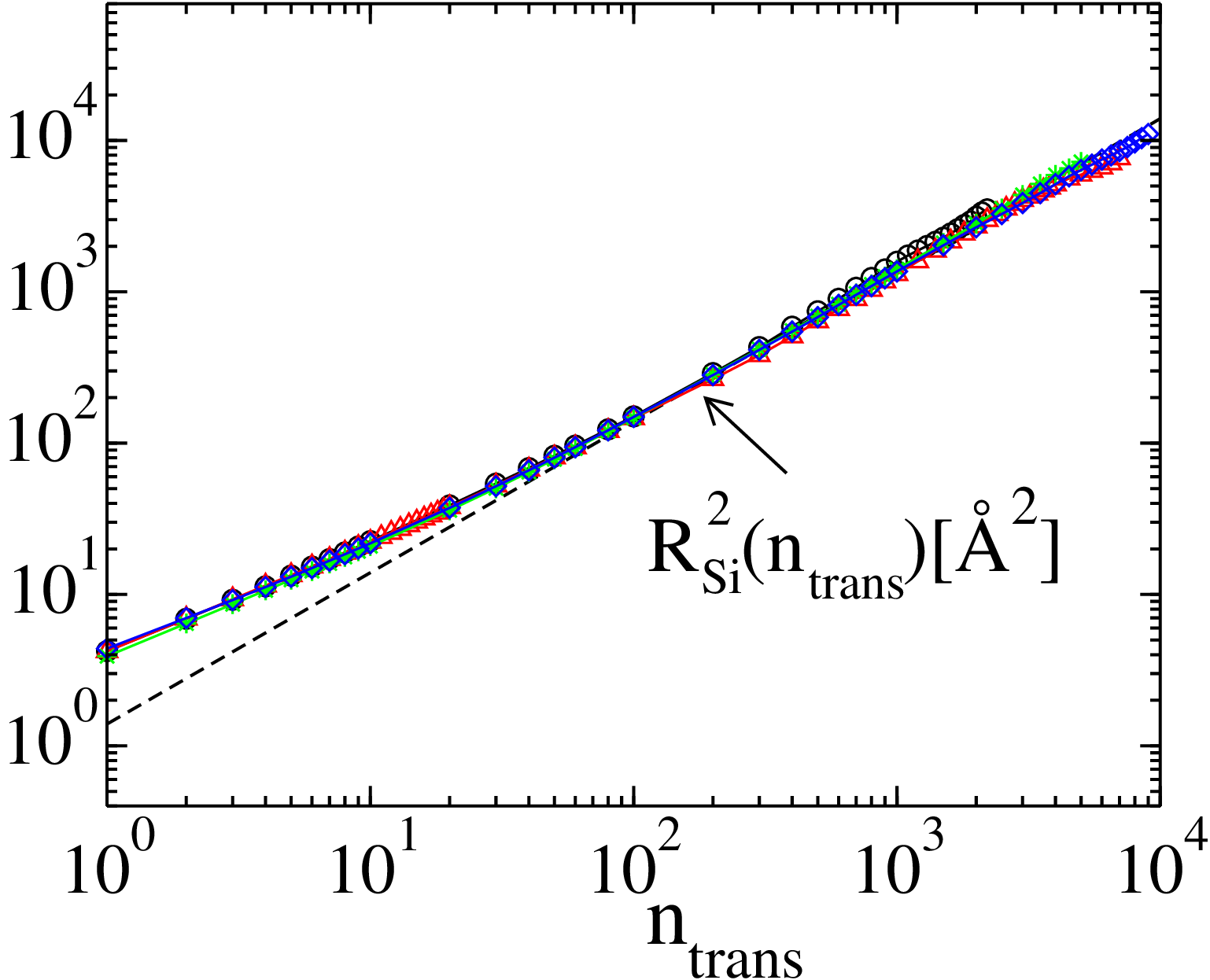}
 \caption{\label{sitransr2}The average mean square displacement of
silicon $R^2_{MB,Si}(n)$ after $n$ MB transitions. In contrast to
Fig.1 only MB transitions within the translational periods are taken
into account.}
\end{figure}

\begin{figure}
\includegraphics[width=8.6cm]{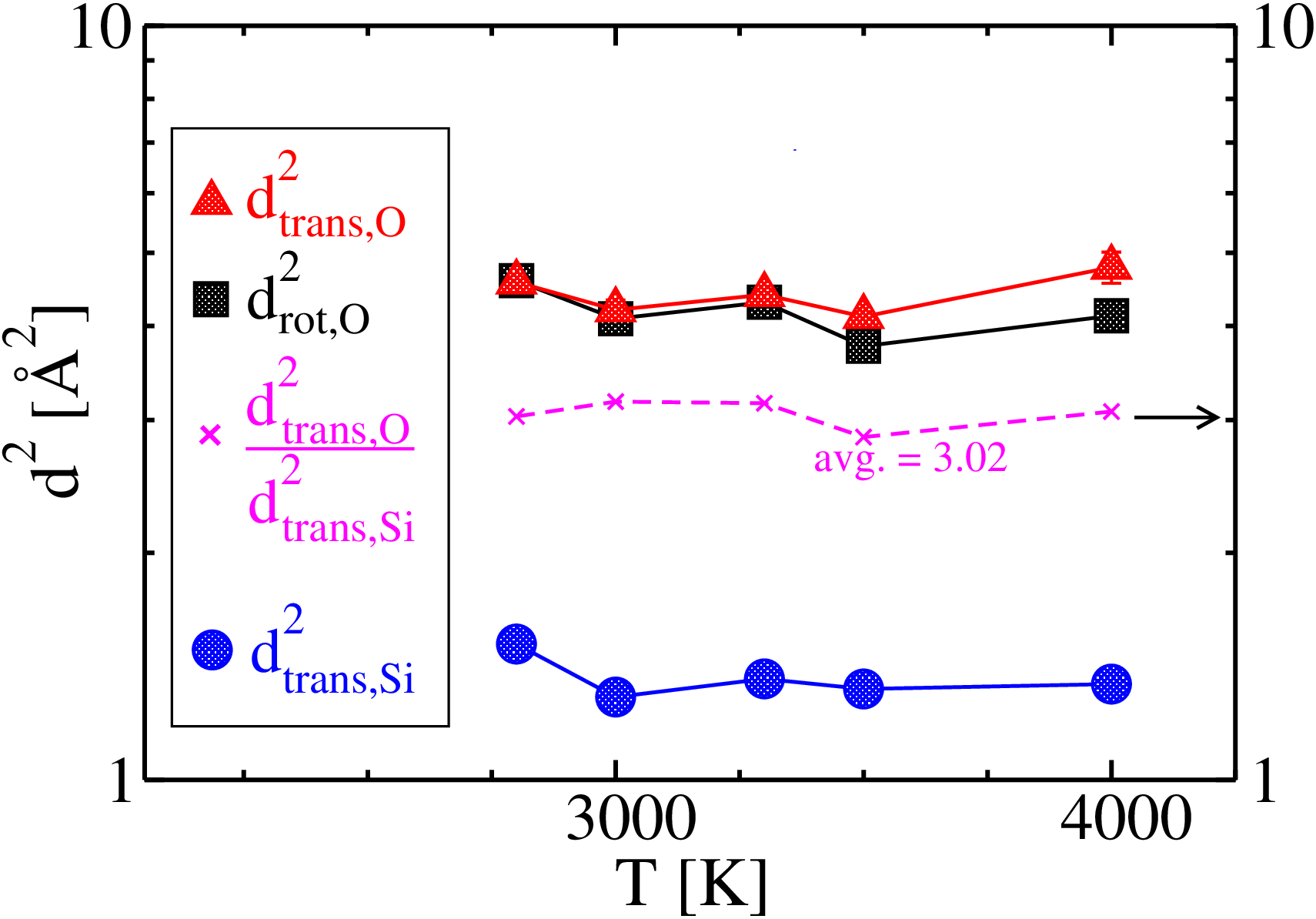}
 \caption{ \label{d2all}Temperature dependence of the rotational and
translational spatial factors $d^2_{trans,O}(T)$,
         $d^2_{trans,Si}(T)$ and $d^2_{rot,O}$ for MB transitions (see definitions in text).}
\end{figure}

The hypothesis that the slowing-down of silicon is exclusively due
to the presence of rotational periods can now easily be tested. For
this purpose we have calculated $R^2_{Si}(n_{trans})$ by ignoring
the MB transitions during the rotational periods. Thus for this
analysis we only count MB transitions during the translational
periods, as indicated by the index {\it trans}. The result is shown
in Fig.\ref{sitransr2}. Within statistical uncertainties one
observes a temperature-independent behavior. This unambiguously
shows that the decoupling of oxygen and silicon dynamics is {\it
exclusively} due to the increasing relevance of rotational processes
at lower temperatures. Furthermore, the non-diffusive part for small
$n_{trans}$, reflecting deviations from a pure random-walk behavior,
becomes smaller and similar in size to that of the oxygen dynamics
in Fig.\ref{osimbr2}. From the large-n$_{trans}$ behavior of
$R^2_{Si}(n_{trans})\propto n_{trans}$ the proportionality constant
can be interpreted as an effective jump-length $d^2_{trans,Si}$. It
is shown in Fig.\ref{d2all}. As expected, no relevant temperature
dependence is observed.

Is the oxygen dynamics different in rotational as compared to
translational periods?  Starting with the oxygen dynamics during
rotational periods we first determine the number of MB transitions
$n_{rot,i}$ and the squared displacement $d_i^2$ during the i-th
rotational period. Formally, $d_i^2$ is the squared Euclidian
distance between the first and the last configuration of a
rotational period.Then one can define the effective oxygen hopping
distance $d^2_{rot,O}$ during rotational periods via
\begin{equation}
\label{eqroto}
 d^2_{rot,O} \equiv \frac{\sum_i
d_i^2}{\sum n_{rot,i}}.
\end{equation}
Its temperature dependence is also shown in Fig.\ref{d2all}. Again,
within statistical uncertainties no temperature dependence is
observed.

Formally, the total displacement of the oxygen atoms during the
translational periods can be obtained from the total mean square
displacement during the whole simulation run minus the contributions
from the rotational periods, i.e. $\sum_i d_i^2$. For the number of
MB transitions $n_{trans}$ in the translational periods one simply
has $n_{trans} = n_{MB} - n_{rot}$. In analogy to Eq.\ref{eqroto}
one can define $d^2_{trans,O}$ as the effective hopping distance
during the translational periods. Again, as seen in Fig.\ref{d2all}
also this observable does not display any significant temperature
dependence.

The ratio of $d^2_{trans,O}/d^2_{trans,Si}$ is close to three which
is larger than the statistical factor of 2. Thus, the silicon
dynamics is indeed somewhat more hindered than the oxygen dynamics.
This reduction in mobility, however, is temperature-independent for
the range of temperatures, analyzed in this work. Apart from the
temperature-independence one observes a surprising similarity of
$d^2_{rot,O}$ and $d^2_{trans,O}$. This implies that, maybe in
contrast to expectation, the nature of the oxygen dynamics in
translational time intervals is similar to that in rotational time
intervals.

\subsection{Transitions between rotational states}

In the final part of this section we discuss some properties of the
transitions between rotational states. These results will shed some
light on the relation between the PEL-properties and the actual
dynamics, as will be discussed in Sect.VI.

\begin{figure}
\includegraphics[width=7.2cm]{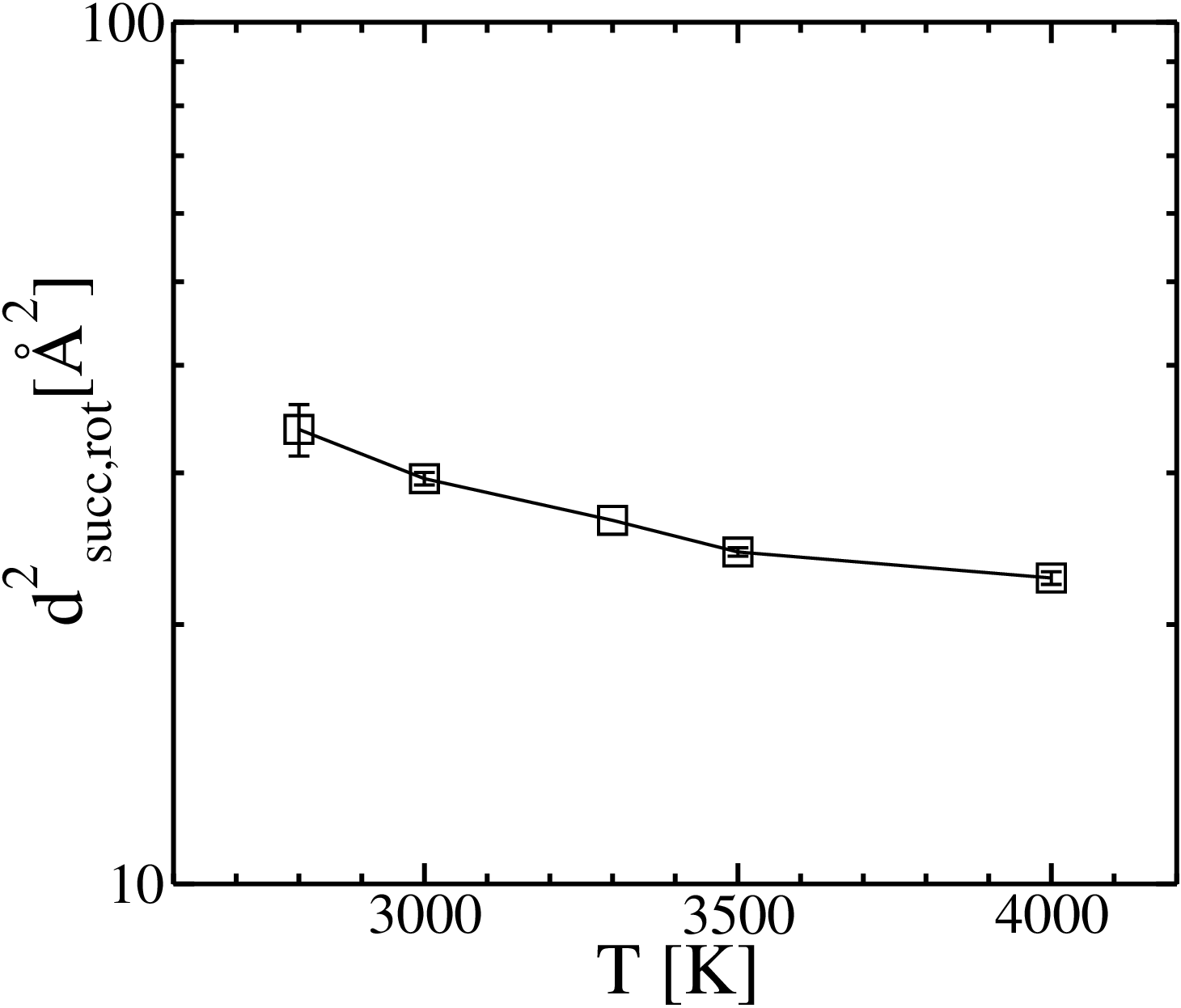}
\caption{\label{d2succ} Temperature dependence of $ d^2_{succ,rot}$
(see definition in text).}
\end{figure}

First, we analyze the typical distance in the configuration space
between successive rotational states. For this purpose we define
\begin{equation}
d^2_{succ,rot,O}\equiv \frac{\sum_i d_i^2}{n_{succ,rot}}
\end{equation}
where $n_{succ,rot}$ denotes the number of transition between
different rotational states, added over all rotational periods. Its
temperature dependence is shown in Fig.\ref{d2succ}. Upon cooling
$d^2_{succ,rot,O}$ increases. This means that the typical distance
between subsequent rotational states becomes larger for lower
temperatures, i.e. more complex oxygen permutations occur. As an
immediate consequence on average more MB transitions are necessary
at lower temperatures to reach a new rotational state within the
same rotational period.

\begin{figure}
\includegraphics[width=7.2cm]{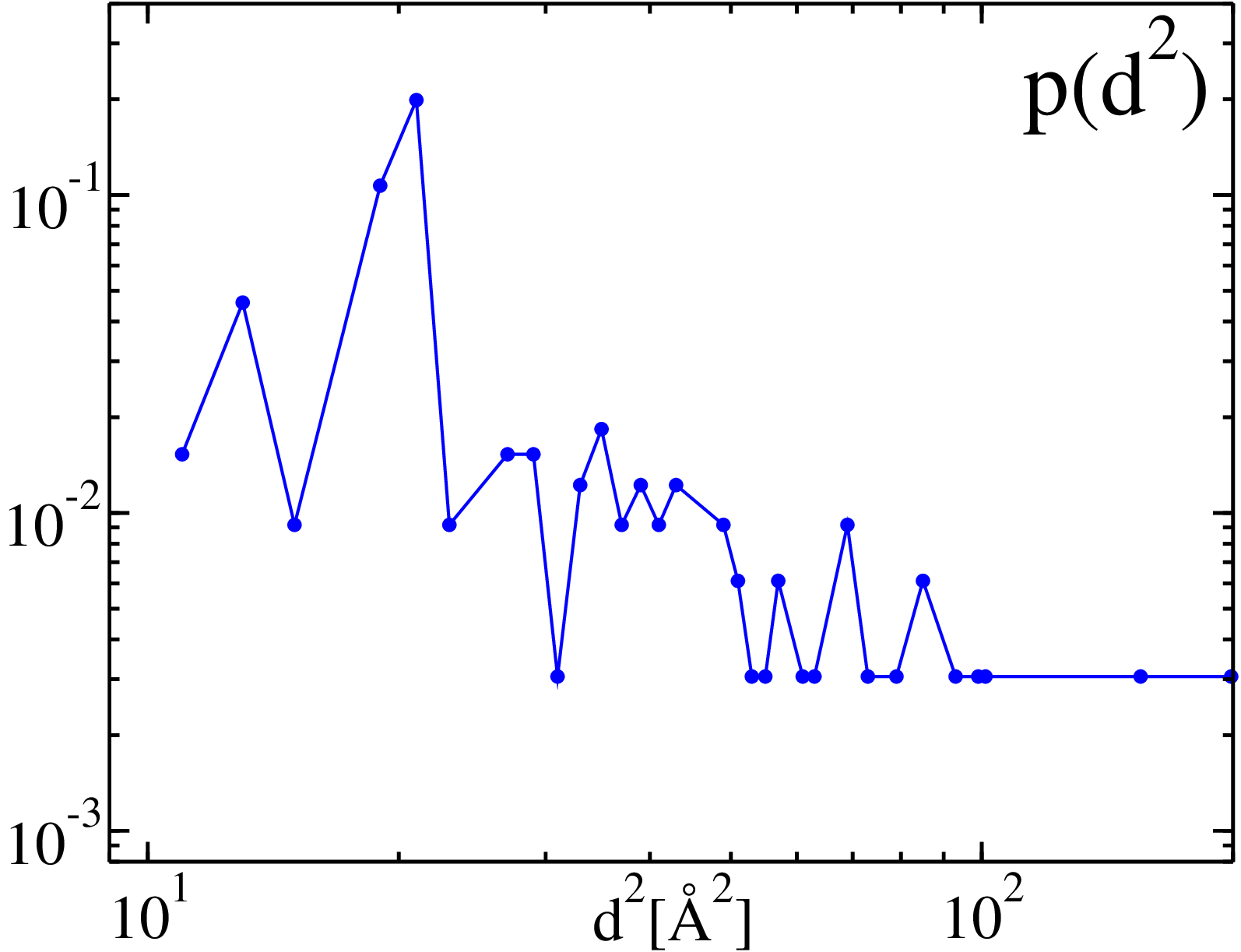}
 \caption{\label{pd2d2}Top:  The distribution of the distances between successive
rotational states at $T = 3000$ K as identified during our MD
simulations. Note the logarithmic scale.}
\end{figure}

More detailed insight into the rotational dynamics can be obtained
by analyzing the distribution of the mean square displacement
between two successive rotational states (here: $T = 3000$ K). This
distribution is shown in Fig.\ref{pd2d2}. One can clearly see that
most displacements are around 20 \AA$^2$ which corresponds to a
simple C$_3$ rotation. A second but significantly smaller peak at
approx. 13 \AA$^2$ reflects the exchange of two adjacent oxygen
atoms. Furthermore, a very small fraction of more complex
permutations with larger displacements is observed.

\section{Quantification of the decoupling of oxygen and silicon dynamics}

\begin{figure}
\includegraphics[width=7.3cm]{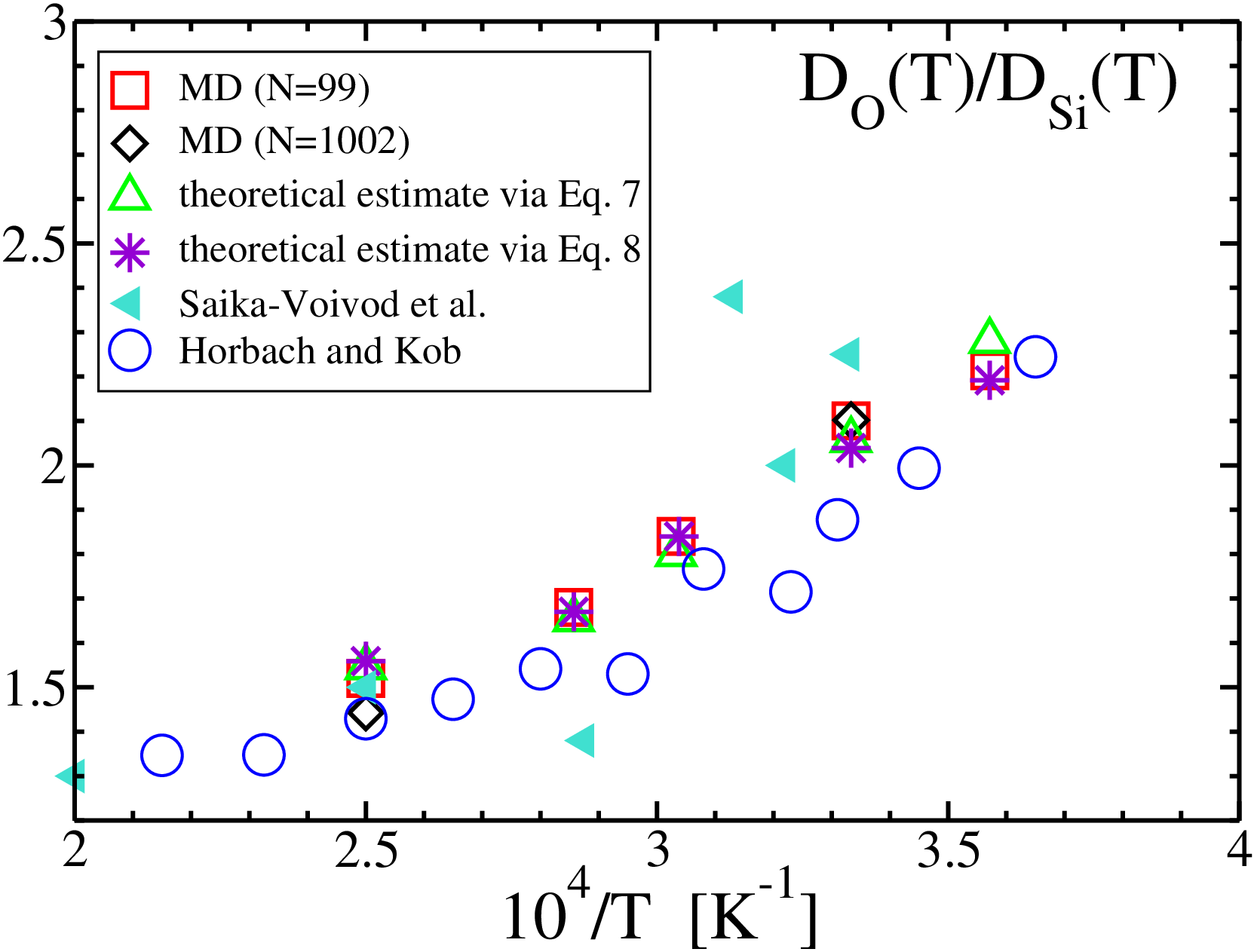}
 \caption{\label{dovsdsi}The decoupling between oxygen and silicon
dynamics.
         Shown is the temperature dependence of
         the ratio of the oxygen diffusion constant to the silicon diffusion constant $D_O/D_{Si}$ as well
          as the theoretical estimations from Eq.7 and
          Eq.8.
         Moreover, the corresponding values from literature \cite{Horbach:1999,Voivod:2004} are included.}
\end{figure}

In this section we first analyze the ratio of the oxygen diffusion
constant to the silicon diffusion constant $D_O(T)/D_{Si}(T)$. Its
temperature dependence, as obtained from our molecular dynamics
simulations, is shown in Fig.\ref{dovsdsi}. Included are also data
from the previous MD work by Horbach et al \cite{Horbach:1999} and
by Saika-Voivod et al \cite{Voivod:2004}. Furthermore we checked for
two temperatures that for our system parameters the ratio does not
depend on the system size within statistical accuracy. Note that the
mild variations of this ratio for different simulations may be due
to the slightly different densities or the different cutoff
conditions for the pair-potentials. In any event, on a
semi-quantitative level all simulations display a very similar
increase of  $D_O(T)/D_{Si}(T)$ with decreasing temperature.

Using the quantities, introduced in the previous section, we may
write
\begin{equation}
\label{eqratio} \mbox{\rule{0mm}{0.3cm}}
\begin{split}
\frac{D_O(T)}{D_{Si}(T)} \ \ &= \ \ \frac{n_{trans}(T) \cdot d^2_{trans,O} + n_{rot}(T) \cdot d^2_{rot,O}}{2 \cdot n_{trans}(T)\cdot d^2_{trans,Si}},\\
\vspace{2mm}
\mbox{\rule{0mm}{0.9cm}}
&=\ \ \frac{d^2_{trans,O}(T)}{2 d^2_{trans,Si}} + \frac{n_{rot}(T) \cdot d^2_{rot,O}}{2 n_{trans} \cdot d^2_{trans,Si}}.\\
\end{split}
\end{equation}
The factor of 2 reflects the stoichiometry of SiO$_2$. Using the
results from Fig.\ref{d2all}, i.e. $ d^2_{trans,O}/d^2_{trans,Si}
\approx 3$ and $d^2_{rot,O} \approx
d^2_{trans,O}$, the ratio $D_O/D_{Si}$ can be rewritten as \\
\begin{equation}
\label{eqratioexplain} \mbox{\rule{0mm}{0.4cm}}
\frac{D_{O}(T)}{D_{Si}(T)} \ \ \approx \ \ \frac{3}{2}\cdot
\left(1+\frac{n_{rot}(T)}{n_{trans}(T)}\right).
\end{equation}
This result quantifies the general statement, formulated above, that
the decoupling is exclusively related to the increasing relevance of
rotations at lower temperatures, as expressed by the temperature
dependence of $n_{rot}(T)/n_{trans}(T)$. The estimated values of
$D_O(T)/D_{Si}(T)$ according to Eq.\ref{eqratioexplain} are also
included in Fig.\ref{dovsdsi}. The excellent agreement with the
simulated $D_O(T)/D_{Si}(T)$ constitutes a consistency check of the
analysis of Sect.III.

\section{Energetic aspects of MB transitions}
MBs are mainly characterized by energy and residence time. In
literature it has been shown that the energy plays an important role
for the dynamics \cite{Saksaengwijit:2006, DoliwaEacc:2003}. Thus,
one may expect that energy also affects the nature of the MB
transitions in the present context. In order to check the energy
dependence for the rotational and translational transitions, we
determine the number of MB transitions in the rotational periods
$n_{rot}(e,T)$ and the translational periods $n_{trans}(e,T)$ in
dependence on MB energy $e$. From Eq.\ref{eqratioexplain}
$D_{Si}(T)/D_O(T)$ can be written as
\begin{equation}
\label{eqener}
\begin{split}
\frac{D_{Si}(T)}{D_O(T)} \hspace{4mm} &= \hspace{4mm}
\frac{2}{3}\frac{ n_{trans}(T)}
             {n_{trans}(T)+ n_{rot}(T)} \\
&= \hspace{4mm} \frac{2}{3} \int de \,  \varphi(e,T) \cdot P_{trans}(e,T)\\
\end{split}
\end{equation}

\begin{figure}
\includegraphics[width=7cm]{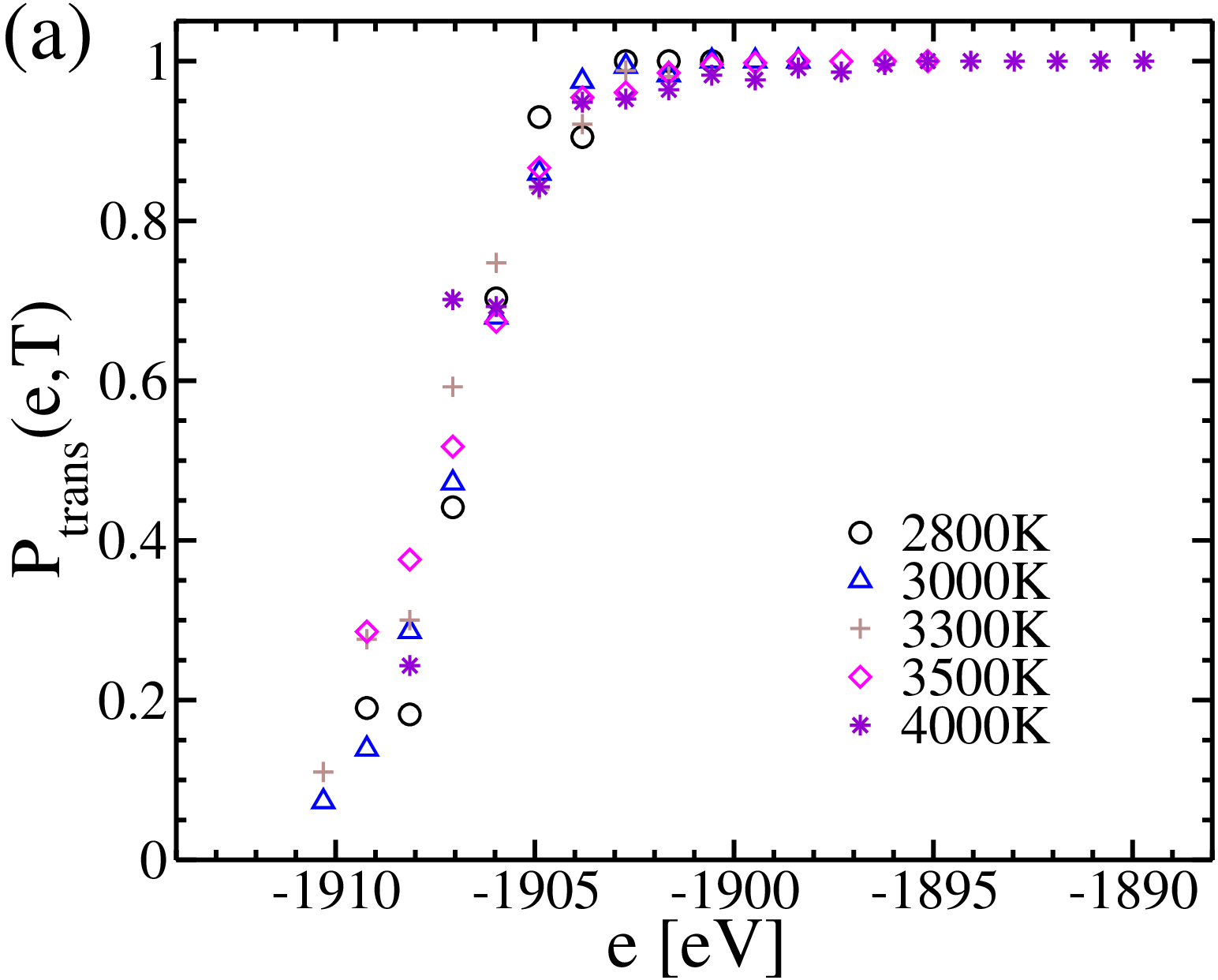}
\includegraphics[width=7cm]{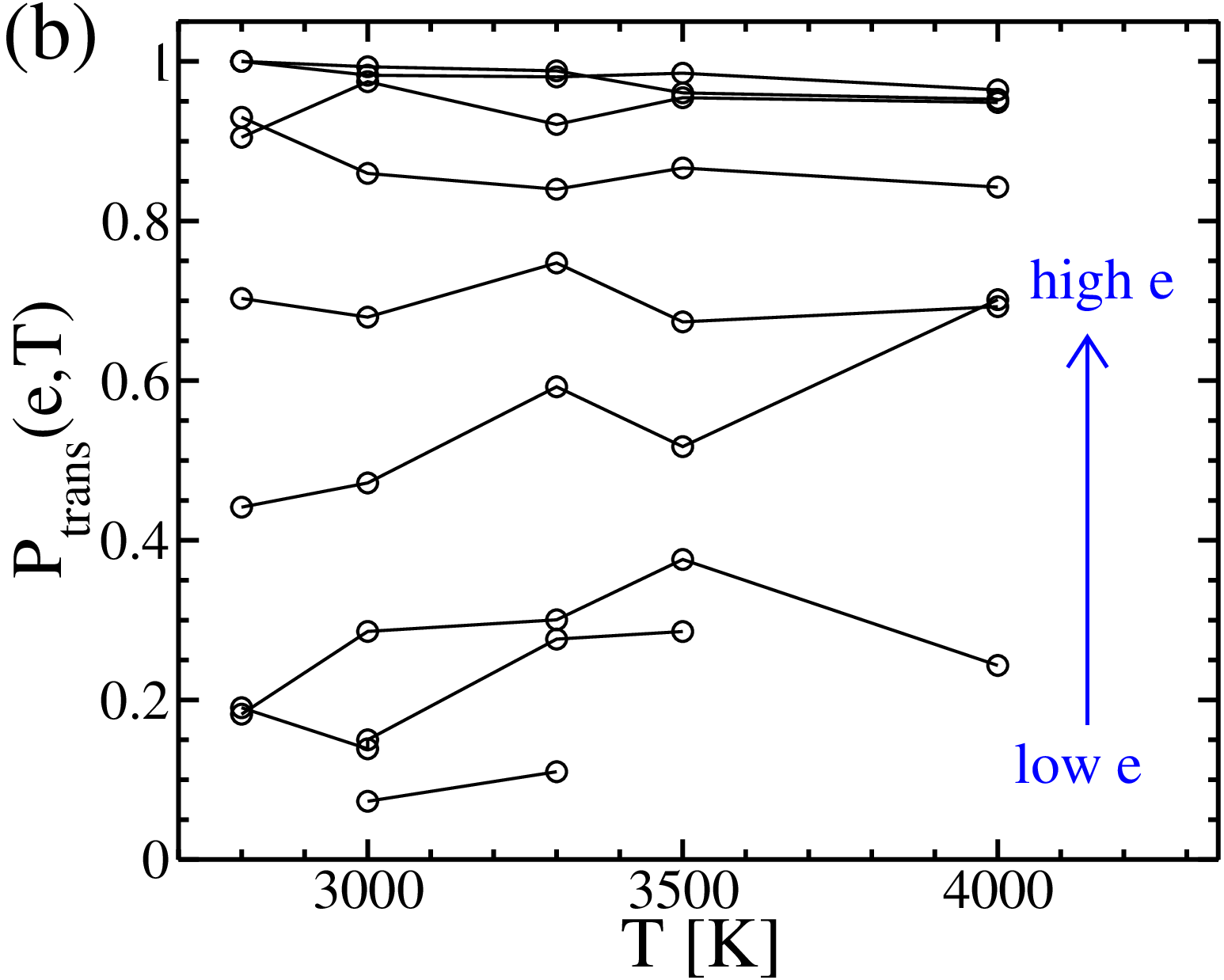}
\caption{\label{ptrans}Energy (a) and temperature (b) dependence of
the probability of translational relaxations
         $P_{trans}(e,T)$.}
\end{figure}

$\varphi(e,T)$ has been defined in Eq.\ref{eqdefvarphi} and denotes
the probability that a MB has the energy $e$. Furthermore,
$P_{trans}(e,T)$ is the probability that a MB of energy $e$ is
visited during a translational period. A small value of
$P_{trans}(e,T)$ thus implies that after leaving a MB of energy $e$
the system very likely will later on end up in a rotational state.
The energy and temperature dependence of $P_{trans}(e,T)$ is shown
in Figs.\ref{ptrans}(a) and (b), respectively. Most importantly,
$P_{trans}(e,T)$ depends strongly on energy whereas it does not show
any significant temperature dependence. In Fig.\ref{ptrans}(a) one
can separate the energy dependence of $P_{trans}(e,T)$ into two
regimes. In the high-energy regime one has $P_{trans}(e,T) \approx
1$. This means that no significant rotational processes happen for
high-energy MB. When the MB energy decreases, rotational processes
become relevant. This results in a decrease of $P_{trans}(e,T)$.
Interestingly, more than 90\% of the transitions involving MB with
energies close to the cutoff energy $e_c$ contribute to the
rotational dynamics. From the present results we have to conclude
that energy not only determines relaxation  rates as shown in the
previous studies \cite{Saksaengwijit:2006, DoliwaEacc:2003} but also
determines the
 relevance of rotational processes.

As a consistent check, we calculate $D_O(T)/D_{Si}(T)$ via Eq.
\ref{eqener}. The results are also included in Fig.\ref{dovsdsi}.
Naturally, a good agreement with the actually observed
$D_O(T)/D_{Si}(T)$ is found.

\section{Discussion and summary}

In previous work on silica we were able to express the oxygen
diffusion constant via
\begin{equation}
D_O(T) \propto \int de \, p(e,T)/\langle \tau(e,T) \rangle.
\end{equation}
where the average waiting time $\langle \tau(e,T) \rangle$ can be
written as
\begin{equation}
\langle \tau(e,T) \rangle = \tau_0 (e) \exp(\beta V(e))
\end{equation}
with an energy-dependent attempt frequency $1/\tau_0 (e)$ and an
effective barrier height $V(e)$.

This set of equations implies that from the three
temperature-independent quantities $G(e),\tau_0(e)$ and $V(e)$ the
thermodynamics as well as the temperature dependence of the oxygen
dynamics (in terms of the diffusion constant) can be predicted. In
particular $V(e)$ can be rationalized in terms of the local topology
of the PEL \cite{Saksaengwijit:2006}. This implies a basically
complete understanding of the oxygen diffusion in terms of
properties of the PEL. In this work we have identified a further
(nearly) temperature-independent function $P_{trans}(e)$ which
allows one to predict the enhanced slowing-down of silicon diffusion
as compared to oxygen diffusion.

\begin{figure}
\includegraphics[width=7cm]{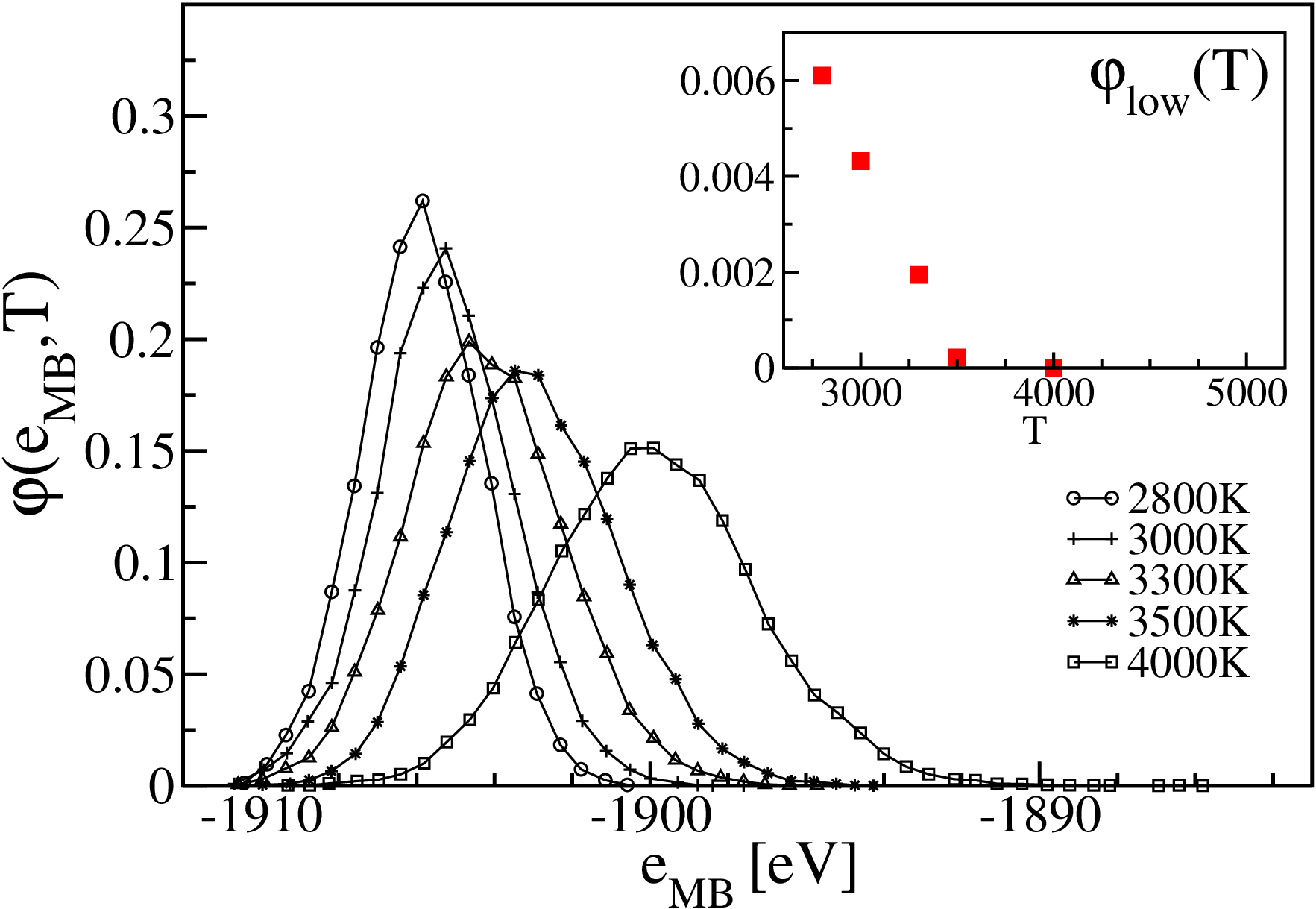}
 \caption{\label{varphi}The distribution $\varphi(e,T)$ of MB for
different temperatures. In the inset the temperature dependence of
$\varphi_{low} = \int_{e_c}^{e_c+1 eV} de \, \varphi(e,T)$ is shown.
It reflects the number of low-energy MB close to the cutoff of the
PEL.}
\end{figure}

In the following we will argue that the results, obtained so far,
are compatible with the general physical scenario of relaxation
processes in silica, as formulated in \cite{Saksaengwijit:2006}. As
a main result it has been shown that there exists a crossover-energy
$ e_{cross} \equiv $ -1905 eV which separates two dynamical regimes.
For $e > e_{cross}$ the escape dynamics is non-activated whereas for
significantly lower $e$ the system first has to climb up in the PEL
to IS with energies close to $e_{cross}$ until it can effectively
escape from this region of configuration space. On a qualitative
level this observation can be related to the strong energy
dependence of the density of MB $G(e)$ or $\varphi(e,T)$,
respectively. In the considered range of temperatures much less than
1\% of all configurations possess an energy less than $e_c + 1$ eV;
see Fig.\ref{varphi}. In contrast, nearly every second MB has an
energy less than $e_{cross}$. As a consequence for MB with $e
> e_{cross}$ the PEL supplies many adjacent configurations of similar energy with
connecting saddles of at most 1 eV which the system may easily
reach, giving rise to non-activated dynamics whereas in the
low-energy limit this is no longer possible due to the lack of such
configurations.

For rationalizing the features, observed in this work, we discuss in
qualitative terms the distribution of MB in configuration space
relative to a given MB (denoted {\it central} MB) with energy $e_0$,
visited at a given temperature $T$. We introduce the function
$\varphi_{all}(e,d^2|e_0)$ as the probability that a randomly
selected MB is a distance $d^2$ away from the central MB and has an
energy $e$. This probability distribution has two contributions
\begin{equation}
\varphi_{all}(e,d^2|e_0) = \varphi_{rot}(e,d^2|e_0) +
\varphi_{rest}(e,d^2|e_0).
\end{equation}

\begin{figure}
\includegraphics[width=7.2cm]{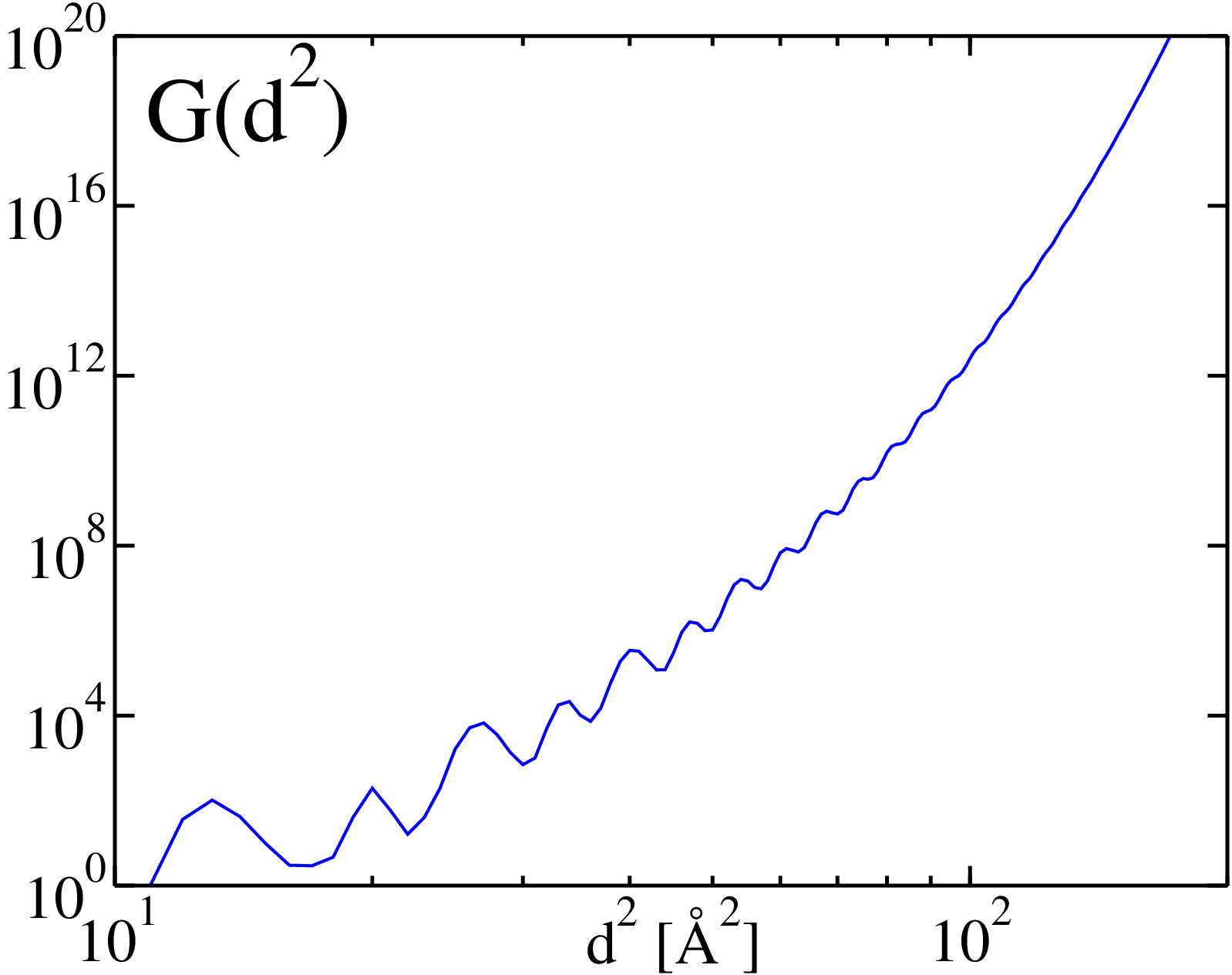}
 \caption{\label{gd2d2} The distribution $G_{rot}(d^2)$ of
all distances between configurations
 with oxygen permutations, as
obtained from Monte-Carlo simulations. }
\end{figure}

The first contribution  takes into account configurations which
result from oxygen permutations of the central MB. This distribution
can be written as $\varphi_{rot}(e,d^2|e_0) \propto G_{rot}(d^2)
\delta (e-e_0)$ where $G_{rot}(d^2)$ is the distribution of
distances, when taking into account {\it all} possible $66!$
permutations. We have calculated  $G_{rot}(d^2)$ for one
representative low-energy configuration and have exchanged the
indices of the oxygen atoms via Monte-Carlo techniques. To determine
the distribution over many orders of magnitude we have used
appropriately adjusted flat-histogram techniques \cite{Shell:2004}.
The resulting distribution $G_{rot}(d^2)$ is shown in
Fig.\ref{gd2d2}. One can clearly see a few peaks for smaller
distances and a transition to a continuous behavior for larger
distances. Naturally, the $d^2$-values of the peaks agree with those
observed during the actual transitions between rotational states.
For $d^2 \ge 100$ \AA$^2$ this distribution no longer shows any
wiggles. This reflects the fact that a transition with a large value
of $d^2$ can be achieved by a multitude of different oxygen
permutations. In contrast, for small $d^2$  discrete distances can
be identified which belong to specific types of reorientations (like
$C_3$ rotations). Comparison with the actually observed transitions
in Fig.\ref{pd2d2} shows that in particular more complex
permutations, related to larger values of $d^2$, are strongly
suppressed. Interestingly, also exchange processes of adjacent
oxygen atoms, corresponding to the first peak, are significantly
suppressed.

The second contribution $\varphi_{rest}(e,d^2|e_0)$ is related to
the remaining MB which have a different structure as the central MB.
For sufficiently large $d^2$ one may naturally write
$\varphi_{rest}(e,d^2|e_0) \propto G_{rest}(d^2) \varphi(e,T)$ with
$\varphi(e,T)$ defined in Eq.\ref{eqdefvarphi}. Qualitatively, this
means that for regions sufficiently far away from the central MB the
distribution of MB is not influenced by the presence of a MB with
$e=e_0$ at the origin. One expects that for large $d^2$ the
distribution $G_{rest}(d^2)$ increases much faster than
$G_{rot}(d^2)$, i.e. with a larger exponent of $d^2$, because
$G_{rest}(d^2)$ contains the exponentially larger number of
different IS as well their permutations (except for those taken into
account by $G_{rot}(d^2)$). Thus, for large $d^2$ one expects
$G_{rot}(d^2) \ll G_{rest}(d^2)$. This dramatic difference may
disappear for small $d^2$ because the ratio of both functions
becomes smaller for smaller $d^2$. Furthermore, in contrast to the
oxygen permutations the possible transitions to adjacent MB are more
continuously distributed with respect to their distances $d^2$. As a
consequence the continuous regime of $G_{rest}(d^2)$ will probably
extend to much smaller values of $d^2$ than observed for
$G_{rot}(d^2)$, i.e. approx. 100 \AA$^2$. Thus, one may speculate
that for small $d^2$ (e.g. $d^2 \approx 20 $ \AA$^2$) $G_{rot}(d^2)$
is of the same order as $G_{rest}(d^2)$. Of course, for a more
quantitative characterization one directly has to determine
$G_{rest}(d^2)$, which, however, is a formidable task.

We note in passing that the factorization of
$\varphi_{rest}(e,d^2|e_0) \propto G_{rest}(d^2) \varphi(e,T)$ will
break down for $d^2 \rightarrow 0$, as known from the study of the
tunneling systems \cite{Reinisch_prb:2004}. There it turned out that
for very nearby IS ($d^2 \le 1 $ \AA$^2$) the energy difference does
not depend on $e_0$ which implies that $\varphi_{rest}(e,d^2|e_0)$
is a function of $e - e_0$.

With the general concepts, introduced above, the following
observations have to be rationalized: (i) The presence of rotations
with small values of $d^2$; see Fig. \ref{pd2d2} but the strong
suppression of rotations with larger $d^2$. (ii) The additional
suppression of particle exchange with $d^2 \approx 13$\AA$^2$. (iii)
The relevance of rotations only for $e < e_{cross}$ and the increase
of rotations for decreasing energy. (iv) The increase of
$d^2_{succ,rot,O}$ with decreasing temperature. (v) The
temperature-independence of $P_{trans}(e)$. (vi)  The similarity of
$d^2_{rot,O}$ and $d^2_{trans,O}$.

(i) For large $d^2$ the function $G_{rest}(d^2)$ will be orders of
magnitude larger than $G_{rot}(d^2)$. Thus for purely statistical
reasons hardly any permutations with large $d^2$ are observed; see
Fig.\ref{pd2d2}. In this context it is surprising that nevertheless
a very small number of permutations with $d^2$ up to 80 \AA$^2$ are
observed which from this purely statistical consideration should be
extremely unlikely. Maybe there are specific pathways on the PEL
which enhance the probability of specific permutations. (ii) For
oxygen exchange processes massive reorientations are necessary in
contrast to, e.g., effective $C_3$ rotations. To perform an exchange
process the system would have to reach regions of the configuration
space which are far away from the central MB. As a consequence, for
purely statistical reasons the probability to complete the exchange
process is very small. (iii) As discussed above for small $d^2$ the
function $G_{rot}(d^2)$ may be of the same order as $G_{rest}(d^2)$.
Furthermore, $\varphi_{rest}(e,d^2|e_0) \propto \varphi(e,T)$ is
very small for $e \approx e_c$, see Fig.\ref{varphi}, implying
$\varphi_{all}(e_0,d^2|e_0) \approx G_{rot}(d^2)
\varphi_{rot}(e_0,d^2)$ for $e_0 \approx e_c$. Thus, for statistical
reasons the relevance of rotational states is significantly enhanced
for very low energies. In contrast, for $e_0 \ge e_{cross}$ the
system possesses a large number of adjacent IS with similar energy
such that again for statistical reasons the probability to reach one
of the rotational states is very small. (iv) As a further
consequence the range of distances for which the probability to find
a rotational state is non-negligible increases with decreasing
energy. Thus, in agreement with observation one would expect an
increase of the average distance between successive rotational
states with decreasing energy and thus with decreasing temperature.
(v) As discussed above for the transition from one low-energy
configuration ($e_0 \approx e_c$) to a new low-energy configuration
the system first has to climb up the PEL to energies close to
$e_{cross}$.  This trajectory basically forms an energy-upwards path
without major intermediate barriers \cite{Saksaengwijit:2006}. In
analogy, the subsequent path to a new low-energy configuration is
not expected to possess significant barriers, in similarity to the
funnel-like picture of protein folding. Stated differently, the
search for the subsequent low-energy configuration is basically
entropy-driven so that $P_{trans}$, indeed, should not significantly
depend on temperature. (vi) Whether or not the system ends up in a
rotational state is mainly a question of the statistical
availability of these states. As a consequence one would expect that
the nature of the oxygen dynamics in translational and rotational
periods should not differ too much. This is consistent with the
observed similarity of $d^2_{rot,O}$ and $d^2_{trans,O}$.

According to Eq.\ref{eqener} the fraction of the rotational and
translational dynamics is related to the thermodynamic weighting of
the MB population, $\varphi(e,T)$. Therefore, the temperature
dependence of $n_{trans}(T)/n_{MB}(T)$ presented above is just a
consequence of the temperature dependence of $\varphi(e,T)$. It is
interesting to estimate the value of $D_O/D_{Si}$ at the glass
transition temperature $T_g$ (1450K). From the density of MBs $G(e)$
and waiting time $\langle \tau(e,T)\rangle $ one can predict
$\varphi(e,T)$ at $T_g$. By using Eq.\ref{eqener} the estimated
value for $D_O/D_{Si}$ at $T_g$ is of the order of $10^1$, which is
one order of magnitude smaller than the experimental value of
$\approx 10^2$ \cite{Mikkelsen:1984, Brebec:1980}. This discrepancy
may be resolved in different ways: (i) At lower temperatures further
decoupling mechanisms set in. (ii) The temperature-independence of
$P_{trans}$  no longer holds for temperatures closer to the
calorimetric glass transition. (iii) The limits of the BKS-potential
to predict the details of the oxygen and silicon dynamics in silica
are reached.

The present PEL approach to characterize the dynamics of oxygen and
silicon allows one to obtain a relatively simple physical picture
about the oxygen dynamics as well as the silicon dynamics. The
essential key ingredient to understand the onset of the decoupling
of oxygen and silicon diffusion is the presence of rotational
dynamics which only contributes to the oxygen dynamics and which can
be rationalized in simple statistical terms in the framework of the
PEL approach. Excluding the rotational periods, the difference in
activation energies of silicon and oxygen disappears in the
considered range of temperatures.

We would like to acknowledge discussions with M. Vogel on this
subject and the NRW Graduate School of Chemistry and the DFG (SFB
458) for the support of this work.


\begin{thebibliography}{30}
\bibitem[{com()}]{comment1}
\bibinfo{note}{In a strict sense $G(e)$ is the density of states weighted by
  the harmonic contributions; see Ref.12 for a closer disucssion. For silica,
  however, this additional contribution is very small.}
\end{thebibliography}

\end{document}